\documentclass[aps,preprintnumbers,amsmath,amssymb,twocolumn,tightenlines,superscriptaddress]{revtex4}

\usepackage{graphicx}
\usepackage{epsfig}
\usepackage{color}

\newcommand{\be}{\begin{eqnarray}}
\newcommand{\ee}{\end{eqnarray}}

 \newcommand{\gsim}{\mathrel{\hbox{\rlap{\lower.55ex \hbox {$\sim$}}
                   \kern-.3em \raise.4ex \hbox{$>$}}}}
\newcommand{\lsim}{\mathrel{\hbox{\rlap{\lower.55ex \hbox {$\sim$}}
                   \kern-.3em \raise.4ex \hbox{$<$}}}}

\newcommand{\ba}{\begin{eqnarray}}
\newcommand{\ea}{\end{eqnarray}}

\begin{document}


\title{Searching for the Subatomic Swirls in the CuCu and CuAu Collisions}
\author{Shuzhe Shi} 
\address{Physics Department and Center for Exploration of Energy and Matter,
Indiana University, 2401 N Milo B. Sampson Lane, Bloomington, IN 47408, USA.}
\author{Kangle Li}
\address{Modern Physics Department, University of Science and Technology of China, No. 96 Jinzhai Road, Hefei, Anhui 230026, China.}
\address{Physics Department and Center for Exploration of Energy and Matter,
Indiana University, 2401 N Milo B. Sampson Lane, Bloomington, IN 47408, USA.}
\author {Jinfeng Liao} \email{liaoji@indiana.edu}
\address{Physics Department and Center for Exploration of Energy and Matter,
Indiana University, 2401 N Milo B. Sampson Lane, Bloomington, IN 47408, USA.}
\address{Institute of Particle Physics and Key Laboratory of Quark \& Lepton Physics (MOE), Central China Normal University, Wuhan, 430079, China.}

\begin{abstract}
Recently the STAR Collaboration discovered the ``subatomic swirls'', that is, the most vortical fluid flow structures in the quark-gluon plasma produced via the AuAu collisions at the Relativistic Heavy Ion Collider (RHIC). Published in Nature and featured as a cover story, this discovery attracted significant interest and generated wide enthusiasm. For such an important finding, it is crucial to look for independent evidences of confirmation and to critically test the current interpretation of the global polarization measurement. We suggest that the CuCu and CuAu colliding systems at RHIC provide such opportunity. Interestingly, our calculations reveal that the fluid vorticity in the CuCu or CuAu collision is comparable to that in the AuAu Collision.   Surprisingly, we find the computed  $\Lambda$ hyperon polarization effect is stronger in the CuCu and CuAu systems than the AuAu system at the same collisional beam energy and centrality class, with an interesting hierarchy  CuCu $>$ CuAu $>$ AuAu due to interplay between $\Lambda$ production timing and the time evolution of the vorticity. 
These predictions can be readily tested by experimental data.
\end{abstract}

\maketitle


\section*{Introduction}

The study of strongly interacting matter under rotation has attracted significant interest recently across disciplines such as condensed matter physics, cold atomic gases, astrophysics and nuclear physics~\cite{Gooth:2017mbd,Fetter:2009zz,2008PhRvA..78a1601U,2009PhRvA..79e3621I,Berti,Watts:2016uzu,Grenier:2015pya,Kharzeev:2015znc,Liao:2014ava}. It has been found that strong fluid rotation can induce anomalous chiral transport effects such as the Chiral Vortical Effect and Chiral Vortical Wave~\cite{Son:2009tf,Kharzeev:2010gr,Jiang:2015cva}. The rotation is also found to influence the phase structures and transitions in various physical systems~\cite{Jiang:2016wvv,Ebihara:2016fwa,Chen:2015hfc,Chernodub:2017ref}. In heavy ion collision experiments, currently carried out at the Relativistic Heavy Ion Collider (RHIC) and the Large Hadron Collider (LHC), it has been long expected that the large angular momentum carried by the colliding system might lead to observable effects such as global polarization of certain produced particles~\cite{Liang:2004ph,Gao:2007bc,Voloshin:2004ha,Betz:2007kg,Becattini:2007sr}.   

Let us focus on the case of heavy ion collisions, in which an extremely hot subatomic material known as a quark-gluon plasma (QGP) is created. The QGP was the form of matter in the early Universe moments after the Big Bang, and is now recreated in laboratory by RHIC and the LHC.  The created hot material has been found to undergo a strong collective expansion that is well described by relativistic hydrodynamics. In a typical non-central collision,  the two opposite-moving nuclei  have their center-of-mass misaligned and thus carry a considerable angular momentum of about $10^{4\sim 5}\hbar$. It has been long believed that a good fraction of this angular momentum will remain in the crated hot QGP  and lead to strong nonzero vortical fluid structures (often quantified by fluid vorticity) during its hydrodynamic evolution, thus forming the ``subatomic swirls''.  Indeed, many  quantitative computational results have suggested their existence~\cite{Becattini:2013vja,Csernai:2013bqa,Csernai:2014ywa,Becattini:2015ska,Jiang:2016woz,Deng:2016gyh,Pang:2016igs,Becattini:2016gvu}, awaiting an experimental verification. 

Recently the STAR Collaboration reported their remarkable discovery of the ``subatomic swirls'' in the AuAu collisions at RHIC, published in Nature and featured as a cover story~\cite{STAR:2017ckg}. A possible signature of the fluid vorticity is the spin polarization of the produced particles which on average should be aligned with the colliding system's global angular momentum direction. But this is an extremely challenging type of measurement. By a clever analysis of spin orientation of the produced subatomic particles called $\Lambda$ hyperons, the STAR Collaboration was able to find very strong evidence for the global polarization effect, from which they extracted an average fluid vorticity of about $10^{21}\, s^{-1}$ , being the most vortical fluid ever known. 

For such an important finding, it is crucial to look for independent evidences of confirmation and to critically test current interpretations of the polarization data. In this paper we propose to use the CuCu and CuAu colliding systems at RHIC as an ideal and natural way to provide the necessary verification. 
Such experiments were performed previously at RHIC and a number of measurements   were previously carried out at several beam energies by PHENIX, PHOBOS and STAR~\cite{Adare:2008ns,Alver:2010ck,Adare:2015cpn,Adamczyk:2016eux}.  
We will compute the vorticity structures in the CuCu and CuAu systems, to be compared with the AuAu system. Surprisingly, our calculations will show that the fluid vorticity in the CuCu or CuAu collision has a similar pattern as and is comparable in magnitude to that in the AuAu Collision. We make quantitative predictions for the polarization measurements versus the collisional beam energy, which can be readily tested by experimental data.

\section*{The Transport Model Setup}

In this study, we use a widely adopted transport model for  heavy ion collisions, the AMPT (A Multi-Phase Transport) model~\cite{Lin:2004en,Lin:2001zk,Lin:2014tya,Ma:2011uma,Shou:2014zsa}. In particular we use the  string melting version of the AMPT model \cite{Lin:2001zk,Lin:2004en} which includes the initial particle production right after the primary collision of the two incoming nuclei, an elastic parton cascade, a quark coalescence model for hadronization, and a hadronic cascade. The AMPT model provides a very reasonable description of the bulk evolution. We will use the same set of parameters as in \cite{Lin:2014tya}, where the simulations with those parameters well reproduced the yields, transverse momentum spectra and $v_2$ data for low-$p_T$ pions and kaons in central and mid-central Au+Au collisions at RHIC. To be specific, the parameters include the Lund string fragmentation parameters ($a=0.55$, $b=0.15$/GeV$^2$), strong coupling constant $\alpha_s=0.33$ for the parton cascade, a parton cross section of 3 mb (i.e. a parton Debye screening mass $\mu=2.265$/fm), and an upper limit of 0.40 on the relative production of strange to nonstrange quarks.

An advantage of the  AMPT model is that it allows explicit tracking of every parton or hadron's motion during the evolution. This allows a relatively straightforward extraction of the system's angular momentum as well as fluid rotation. The model was first used in \cite{Jiang:2016woz} to compute the structures of local fluid vorticity:
 \begin{eqnarray}
 \Omega_{\mu\nu}=\frac{1}{2}(\partial_\nu u_\mu - \partial_\mu u_\nu)
 \end{eqnarray}
 where $u^\mu$ is the four-velocity of fluid cells computed in AMPT  from averaging  thousands of events for a given collision energy and impact parameter. The interesting component is the one along the out-of-plane direction $\omega_y \equiv \Omega_{3,1}$, which in the non-relativistic limit becomes the familiar y-component of the three-vorticity $\vec{\bf\omega} = \frac{1}{2} \nabla \times \vec{\bf v}$
with $\vec{\bf v}$  the three-velocity.  The global rotation effect can be quantified by properly averaging over the whole fireball. The results for AuAu collisions obtained in  \cite{Jiang:2016woz} predicted a strong monotonic decrease of the average vorticity with increasing beam energy, in consistency with later experimental data~\cite{STAR:2017ckg}. Another advantage of the AMPT model is that the finally observed hadrons are explicitly formed via quark coalescence. It is relatively easy to incorporate the spin polarization effect upon the formation of hadrons such as the $\Lambda$ hyperon. More specifically, the spin 4-vector of each $\Lambda$ is determined from the local vorticity at its formation location, as~\cite{STAR:2017ckg,Becattini:2016gvu,Li:2017slc}: 
\begin{eqnarray}
S^\mu\equiv-\frac{1}{8mT}\epsilon^{\mu\nu\rho\sigma}p_\nu \Omega_{\rho\sigma}
\end{eqnarray} 
where $p^\nu$ is the four-momentum of the hyperon and $m$ the hyperon mass. 
Such polarization could then be properly converted to the $P_\Lambda$ observable measured by the STAR in \cite{STAR:2017ckg}. A subsequent AMPT study \cite{Li:2017slc} for AuAu collisions included the polarization due to local vorticity for each produced hadron in the AMPT framework to calculate the overall spin polarization projected onto the global angular momentum direction. The results for $\Lambda$ polarization are in agreement with the STAR data.

\begin{figure}[!hbt]
\begin{center}
\includegraphics[scale=0.36]{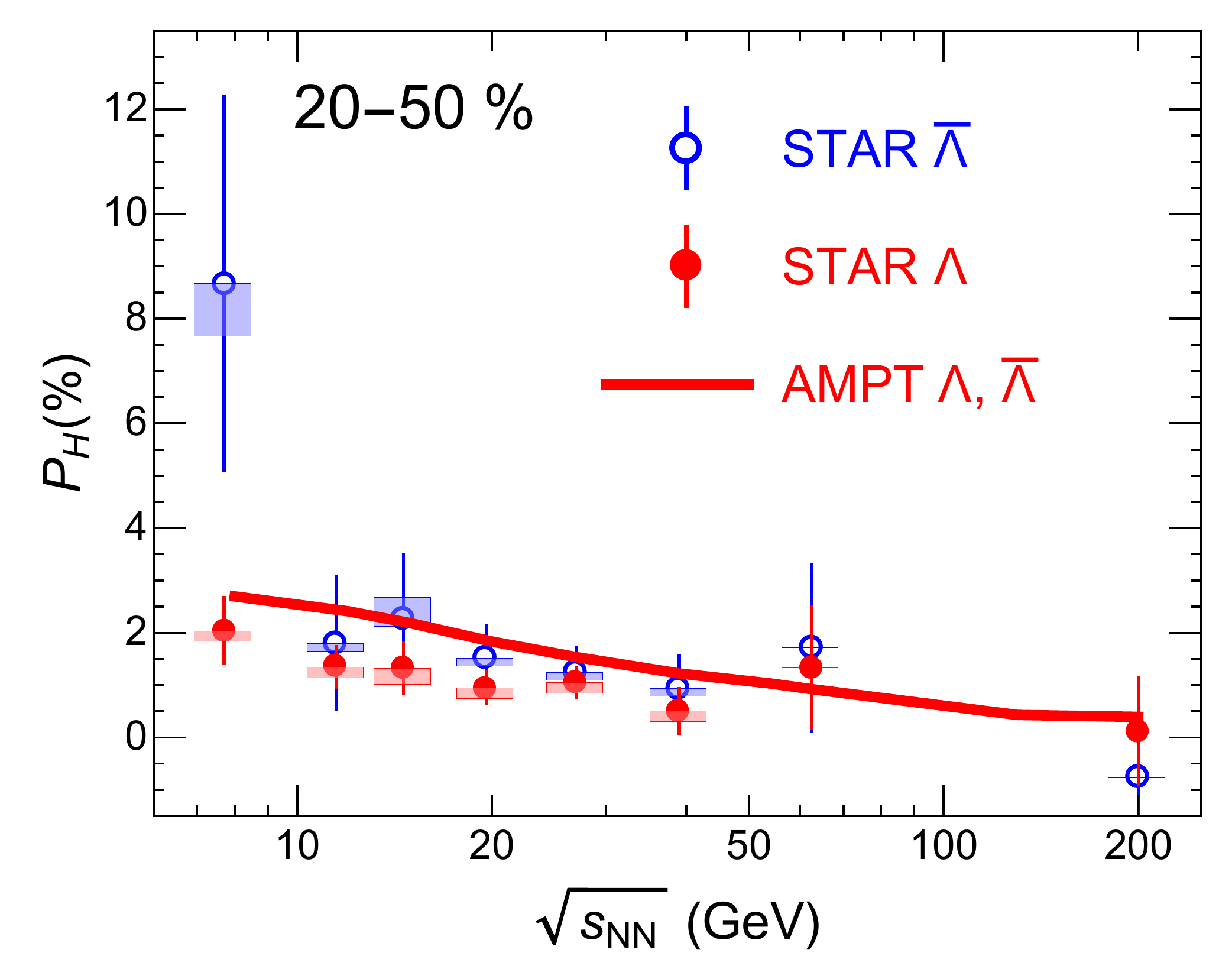}
\caption{(color online)  The $\Lambda$ hyperon global polarization $P_H$ results for AuAu collisions at RHIC, computed from the AMPT model (see text) and compared with the STAR data.}
\label{fig1}
\end{center}
\end{figure}

A versatile aspect of the present model is that it allows easy adaptation to be used for studying various different colliding systems across a wide beam energy span. In this study, we use the same AMPT model setup and computing methods as that in \cite{Jiang:2016woz}, albeit for the investigations of two different  systems: the CuCu and CuAu collisions.   Our goal is to compute the fluid vorticity in these systems and to make quantitative predictions for the polarization measurements versus the collisional beam energy, which can be readily tested by future experimental analysis. As a step of model validation, we first use the AMPT to compute the $\Lambda$ polarization in AuAu collisions. Note that in the calculation we use the same centrality class and kinematic selection as the STAR measurement. Furthermore based on the estimates by STAR~\cite{STAR:2017ckg}, we've taken into account the 20\% suppression effect on top of the primary $\Lambda$ polarization directly computed from the model due to the hyperons from resonance decays. The results, shown in Fig.~\ref{fig1}, are consistent with results from \cite{Li:2017slc} and compare well with the STAR data.  This demonstrates that  predictions from this approach are quantitatively robust.

\section*{Fluid Vorticity and Particle Polarization for CuCu and CuAu Collisions}

\begin{figure*}[!hbt]
\begin{center}
\includegraphics[scale=0.38]{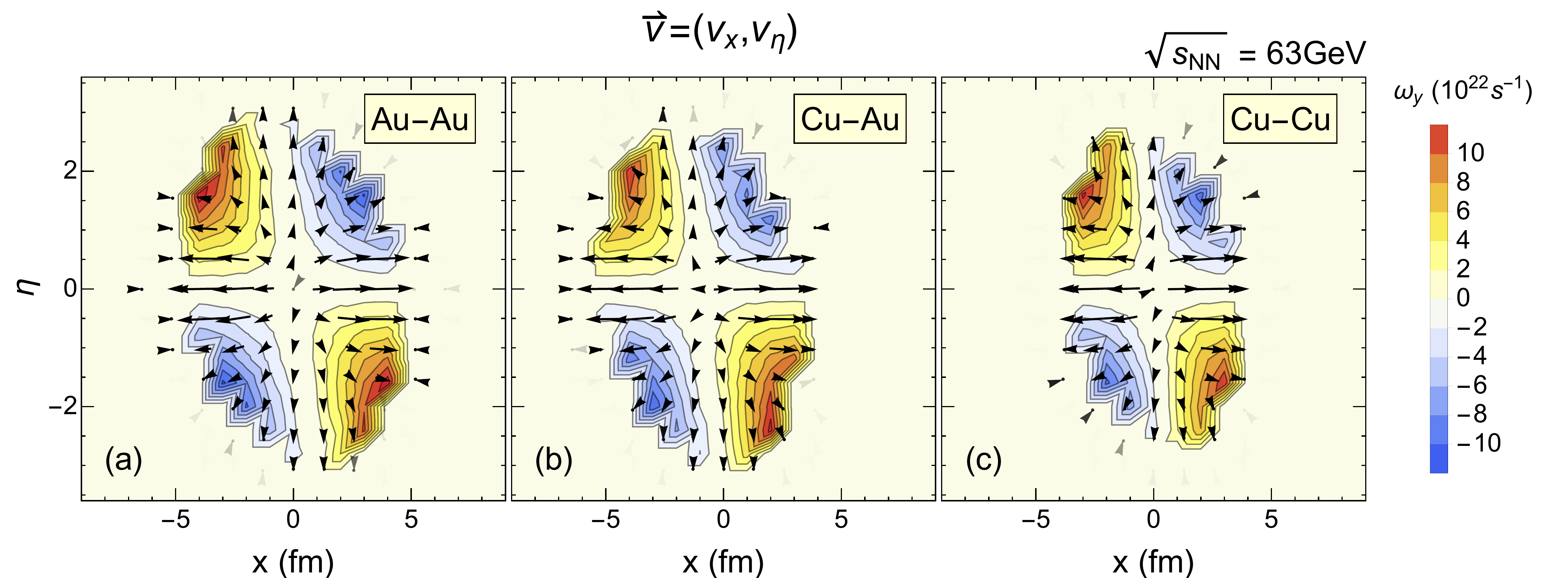}
\caption{(color online) The velocity component distribution $\vec{v}=(v_x,v_\eta)$ (black arrows) on the $x$-$\eta$ plane  and  the corresponding vorticity component $\omega_y$ distribution (colored contour plot), for 20-50\% AuAu, CuAu and CuCu collisions  at $\sqrt{s_{NN}}=63\rm GeV$ at time $t=2$ fm/c. }
\label{fig2}
\end{center}
\end{figure*}

Let us know present the new results for CuCu and CuAu collisions, in comparison with AuAu collisions. As already demonstrated in previous works~\cite{Jiang:2016woz,Deng:2016gyh,Pang:2016igs}, the main contributions to the $\omega_y$ arise from the nontrivial distribution distribution patterns of the fluid velocity vectors on the reaction plane i.e. the $x$-$z$ or $x$-$\eta$ plane. To give an intuitive picture of the vorticity structures in these different systems, we first show in Fig.~\ref{fig2} the velocity component distribution $\vec{v}=(v_x,v_\eta)$ on the $x$-$\eta$ plane as well as  the corresponding vorticity component $\omega_y$ distribution. The three panels are for  AuAu, CuAu and CuCu collisions respectively, all computed for 20-50\% centrality at $\sqrt{s}=63\rm GeV$ at time $t=2$fm/c. While we've done calculations at a variety of beam energies, we choose this particular energy for comparison as  experimental data were taken at this energy for all three systems. From Fig.~\ref{fig2}, one sees that the vorticity distribution patterns, consistent with previous AuAu results~\cite{Jiang:2016woz}, are highly similar among these different systems. What appears somewhat surprising is that, despite the difference in system size and spatial spread of fireball,  even the absolute magnitude of vorticity component $\omega_y$ is  very close for all of them. We've verified this observation to be true for  calculations at varied values of the collisional beam energy.

\begin{figure}[!hbt]
\begin{center}
\includegraphics[scale=0.35]{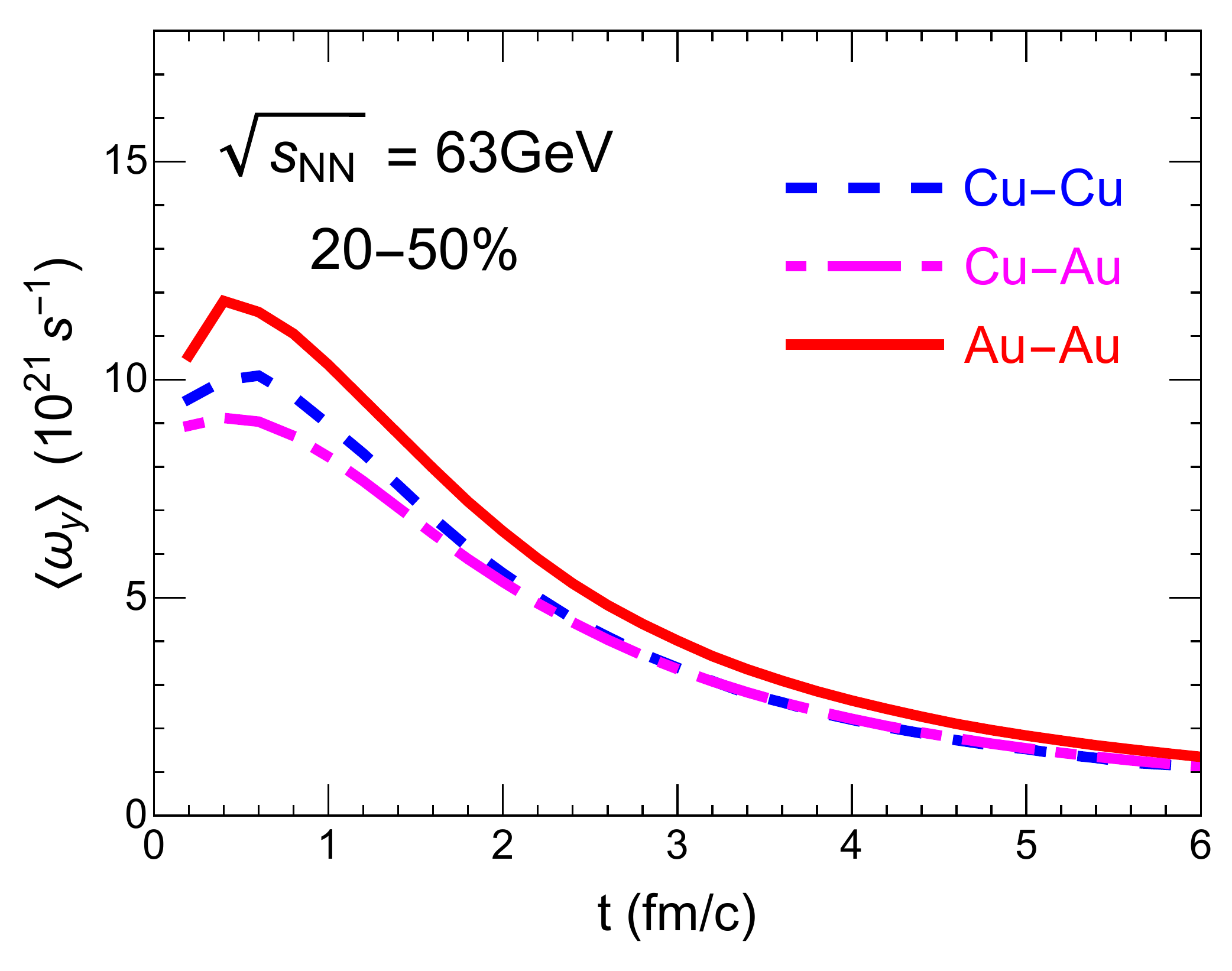}
\vspace{-0.1in}
\caption{(color online) The averaged vorticity $\langle\omega_y\rangle$ as a function of time, for 20-50\% AuAu, CuAu and CuCu collisions  at $\sqrt{s_{NN}}=63\rm GeV$ respectively.}
\vspace{-0.15in}
\label{fig3}
\end{center}
\end{figure}

In order to more quantitatively compare the systems, we compute the averaged vorticity component $\omega_y$ as a function of time. While the local vorticity $\omega_y$ varies a lot and even oscillates in sign across the fireball (as seen in Fig.\ref{fig2}), after properly averaging over the fireball,  the obtained $\langle\omega_y\rangle$ has a definitive sign that aligns with the global angular momentum and is directly relevant for  the size of the observable polarization effect in the end. We use the same averaging procedure as that used for the AuAu calculations in \cite{Jiang:2016woz} to compute the $\langle\omega_y\rangle$ for the CuAu and CuCu systems. The results are shown in Fig.~\ref{fig3}.  Indeed the values of  $\langle\omega_y\rangle$ are found to be fairly comparable for all of them with slightly larger values for the AuAu system, confirming the qualitative observation from Fig.\ref{fig2}. This quantitative comparison also implies that the potentially observable signals, i.e. the particle polarization effects, should also be comparable thus equally measurable for CuAu and CuCu systems as well.

\begin{figure}[!hbt]
\begin{center}
\includegraphics[scale=0.35]{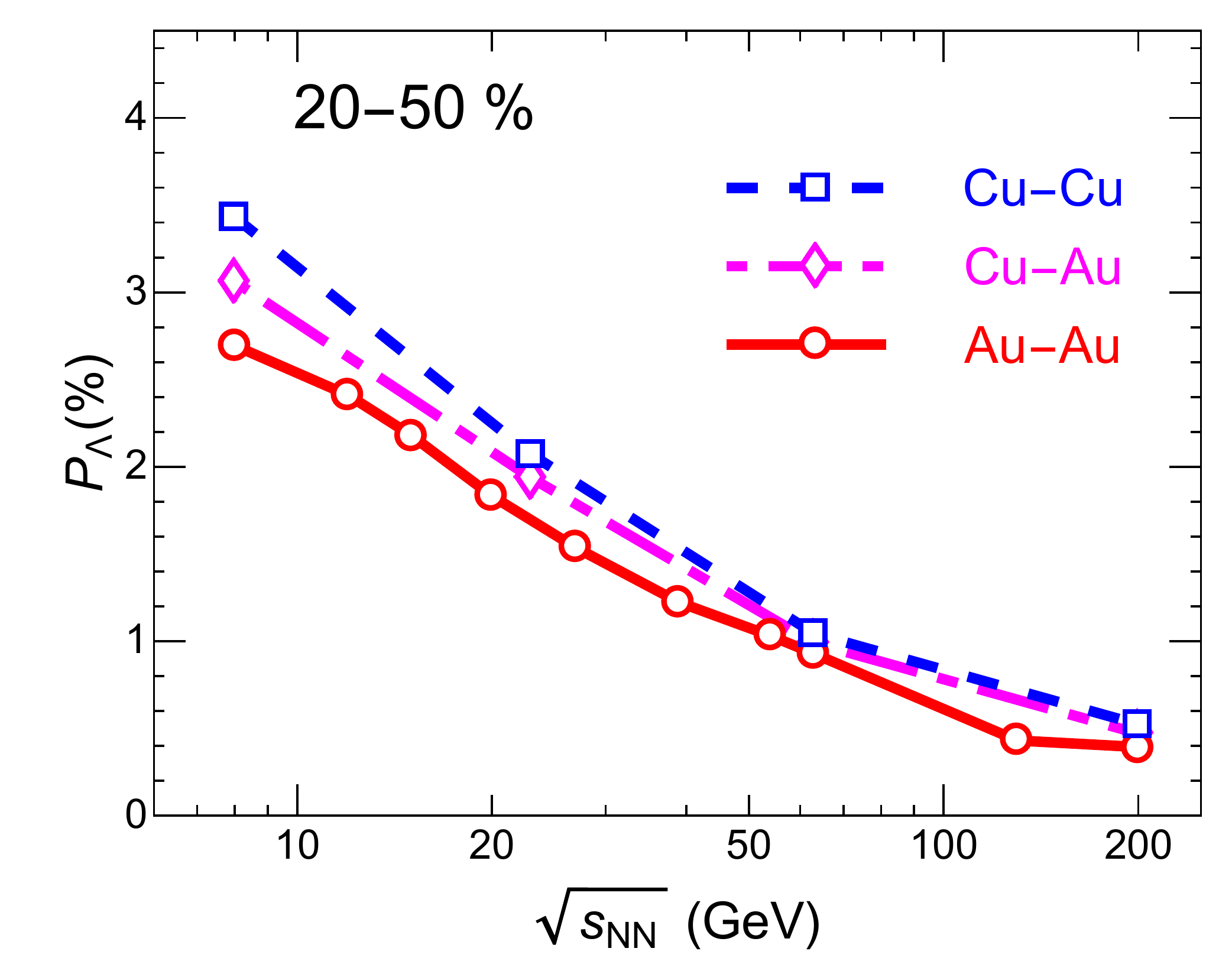}
\vspace{-0.1in}
\caption{(color online) The predicted $P_\Lambda$ signals versus collisional beam energy $\sqrt{s_{NN}}$ for  20-50\% CuCu and CuAu collisions in comparison with the AuAu collisions.}
\vspace{-0.15in}
\label{fig4}
\end{center}
\end{figure}

To verify such expectation, we then compute the $\Lambda$ hyperon polarization, in the same way as the calculation for AuAu (which is validated with data in Fig.\ref{fig1}). The results for the $P_\Lambda$ versus collisional beam energy $\sqrt{s_{NN}}$ are shown in Fig.\ref{fig4} for 20-50\% AuAu, CuAu and CuCu collisions respectively. To our surprise, while the predicted signals are indeed comparable, we find a very interesting hierarchy of the signal strength: CuCu $>$ CuAu $>$ AuAu. At first sight, this appears counterintuitive, especially given that the averaged vorticity is slightly bigger in AuAu system as shown in Fig.~\ref{fig3}. Upon careful examination, it is found that the key to understand this hierarchy is the timing of the hyperon production. Because CuCu system is much diluter than the CuAu system which is still diluter than the AuAu system, the fireball evolution time till freeze-out is generally shorter. As a result, the $\Lambda$ hyperons are on average formed earlier in CuCu system while later in AuAu system. This statement has been explicitly checked in our numerical calculations: the $\Lambda$ production rate curve versus evolution time is visibly shifted to the earlier time side than that in the AuAu collisions. This timing effect, put in the perspective of the rapid decrease of $\langle\omega_y\rangle$ with time in a similar fashion for all three systems (shown in Fig.~\ref{fig3}), implies that the $\Lambda$ produced in CuCu system is on average influenced by a larger vorticity value thus shows stronger polarization effect, as compared with that in AuAu system. Our predictions for the hyperon polarization in CuCu and CuAu systems and for the interesting hierarchy in the signal strength could be readily tested with experimental analyses.

\section*{Summary}

In summary, we've used a transport model to investigate the fluid vorticity structures as well as the global hyperon polarization effect in two new colliding systems: the CuCu and CuAu collisions. A detailed picture of the vorticity structures in those systems was obtained and found to be very similar to that in AuAu collisions at the same beam energy and centrality class. The  average vorticity component along the out-of-plane direction was quantitatively extracted and also found to be comparable with and has a similar time-dependence as that in AuAu collisions. A most unexpected finding comes from the computed final state hadron observable, namely the $\Lambda$ hyperon polarization, where a stronger signal is found in the CuCu and CuAu systems than the AuAu system. The interesting hierarchy of polarization signal (CuCu $>$ CuAu $>$ AuAu) could be understood from the time dependence of vorticity evolution and the different timing for the hyperon production in these systems. These predictions can be readily tested by experimental data, and we propose the global hyperon polarization measurements in CuCu and CuAu collisions as an ideal and independent verification for the subatomic swirl discovery and for our current interpretation of the observed signal as rotational polarization effect.   

Let us mention in passing that we've also studied another potential observable: the  $\phi$ meson polarization for these colliding systems over a wide beam energy range. Naively one expect a stronger polarization effect on $\phi$ which is a spin-1 vector meson. However experimentally it is not possible to separately decipher the probability over all three polarization states $S_{\hat{p}}=\pm1,0$, and what could be measured is only its deviation from isotropic situation. This is to be quantified by the deviation  of its probability on $S_{\hat{p}}=0$ state, i.e. $(1-3\rho_{00})$, which only picks up a quadratic correction from  rotational polarization $\sim (\omega_y/T)^2$. (In contrast, the measured hyperon polarization is a linear order effect.)  Our calculations find the measurable $\phi$ polarization signal to also show a similar hierarchical pattern of CuCu $>$ CuAu $>$ AuAu, albeit all with rather small magnitude, on the order of   $(2\sim5)\times 10^{-3}$.
  
Last but not least, it would be of great interest to explore the vorticity-driven anomalous chiral transport effects in all these systems, once the fluid vorticity structures are well established.  Similarly to the efforts for studying chiral magnetic effects~\cite{Guo:2017jxs,Jiang:2016wve}, the quantitative study of chiral vortical effects would require the future development of hydrodynamic frameworks that incorporate both the 3D fluid vorticity structures and the anomalous transport currents required by chiral anomaly.  Using the different colliding systems for contrast may provide useful insights especially given the difficulty of background contamination: from CuCu to CuAu to AuAu, the vorticity would be comparable while the bulk background contributions may differ considerably.

 \vspace{0.1in}

 {\bf Acknowledgments.} 
The authors thank Yin Jiang, Hui Li, Ziwei Lin and Xiaoliang Xia  for helpful discussions.   This material is based upon work supported by the U.S. Department of Energy, Office of Science, Office of Nuclear Physics, within the framework of the Beam Energy Scan Theory (BEST) Topical Collaboration. The work is also supported in part by the National Science Foundation under Grant No. PHY-1352368 (SS and JL) and by the National Science Foundation of China under Grant No. 11735007 (JL). KL acknowledges the support of the University of Science and Technology of China via an exchange student program scholarship.

\end{document}